\documentstyle[12pt]{article}
\begin{document}

\vspace{0.5cm}

\begin{center}
{\bf{INDUCED RIGID STRING ACTION FROM FERMIONS}}
\end{center}

\vspace{1.0cm}

\begin{center}
R.Parthasarathy{\footnote{e-mail address:
sarathy@imsc.ernet.in}} \\
The Institute of Mathematical Sciences \\
Chennai-600 113, India. \\
and \\
K.S.Viswanathan{\footnote{e-mail address: kviswana@sfu.ca}}
\\
Department of Physics \\
Simon Fraser University \\
Burnaby. B.C. \\
Canada V5A 1S6. \\
\end{center}

\vspace{1.0cm}

{\noindent{\it Abstract}}

\vspace{0.5cm}

From the Dirac action on the world sheet, an effective action
is obtained by integrating over the 4-dimensional  
fermion fields
pulled back to the world sheet. 
This
action consists of the Nambu-Goto area term with right
dimensionful constant in front, extrinsic curvature action and
the topological Euler characteristic term. 

\vspace{1.5cm}

{\bf 1.} QCD strings, in contrast to the fundamental strings,
are described by an action
\begin{eqnarray}
S &=& T \int \sqrt{g}\  d^2\xi + \alpha \int \sqrt{g}\ {\mid H
\mid}^2\ d^2xi,
\end{eqnarray}
where $T$ is the string tension (of dimension
${\it{length}}^{-2}$), $\alpha$ is a dimensionless constant -
measure of the stiffness of QCD flux tubes , $H^i\ =\
\frac{1}{2} g^{\alpha\beta}H^i_{\alpha\beta}$; and  $\ {\mid
H\mid}^2\ =\ H^iH^i$ with sum over $i$, the extrinsic
curvature of the string world sheet. Here and in what follows
$g_{\alpha\beta}$ is the induced metric on the world sheet
which will be regarded as an immersion in $R^4$. The above
action (1) has been proposed for describing QCD strings by
Polyakov [1] and independently by Kleinert [2]. By considering
the second term alone, it has been shown [1,2,3] that the
dimensionless coupling $\alpha$ is asymptotically free.
Further, by evaluating the 1-loop effective action for minimal
and harmonic immersions in the instanton background [4], the
partition function for the action (1) has been shown to be
that of a modified 2-dimensional Coulomb system. Invoking the
running coupling constant ${\alpha}_R$, the system at low
momentum region prefers to be in a phase with long range
correlations, thereby suggesting confinement of the flux
tubes. Recently, Kleinert and Chervyakov [5] have argued
comparing  the high temperature behaviour of QCD string (1)
with that of large-N QCD results of Polchinski [6] that the
stiffness parameter $\alpha$ has to be negative.

\vspace{1.0cm}

{\bf 2.} The action (1) thus appears promising to describe QCD
strings in the non-critical dimensions and a derivation of (1)
will throw more insight into the issues involved. Wiegmann [7]
has demonstrated first that the action of spinning string
requires the {\it extrinsic} geometry of the embedding of the
world sheet into the flat background and later [8] showed that
such an action can be derived from the covariant action of
Neveu-Schwarz-Ramond and Green-Schwarz {\it {super strings}}
by the integration over the fermionic fields. Polyakov [9] has
considered the expectation value of Wilson loops in compact
$U(1)$ gauge theory in the presence of monopoles and obtained
the action (1) in the large loop limit. It will be important
to examine the non-supersymmetric analogue of [8] without
recourse to string actions and the
present work is devoted to this issue. This is along the lines
of and inspired by the work of Rajasekaran and Srinivasan [10]
who showed by integrating over the fermion fields in the Dirac
action $\int d^4x \bar{\psi}\{{\partial}\!\!\!/ +
{A}\!\!\!/ \}\psi$, where $A_{\mu}\ =\ A^a_{\mu}{\lambda}^a$ the
gluon field, one obtains the full lagrangian for QCD. The
field strength $F^a_{\mu\nu}$ is induced by the fermion
integration.

\vspace{1.0cm}

{\bf 3.} The Dirac action induced on a 2-dimensional surface
$\Sigma$ immersed in $R^4$ is 
\begin{eqnarray}
\int {\cal{L}}_D  d^2\xi &=& \int_{\Sigma} d^2\xi \ \sqrt{g}
\frac{i}{2}
g^{\alpha\beta}\bar{q}\{{\gamma}_{\alpha}
{\partial}_{\beta}\!\!\!\!\!^{^{^{\rightarrow}}}
 - {\partial}_{\beta}\!\!\!\!\!^{^{^{\leftarrow}}}{\gamma}_{\alpha}\}q,
\end{eqnarray}
where the Dirac spinor $q$ is in $R^4$ and ${\gamma}_{\alpha}$ is
the pull-back of the euclidean Dirac gamma matrices
${\gamma}_{\mu}$ ($\mu$ = 1,2,3,4)
\begin{eqnarray}
{\gamma}_{\alpha} &=& {\partial}_{\alpha}X^{\mu}\
{\gamma}_{\mu}.
\end{eqnarray}
Using,
\begin{eqnarray}
\{{\gamma}_{\mu}, {\gamma}_{\nu}\} &=& 2{\delta}_{\mu\nu}
\nonumber
\end{eqnarray}
it follows
\begin{eqnarray}
\{{\gamma}_{\alpha}, {\gamma}_{\beta}\} &=& 2g_{\alpha\beta}.
\end{eqnarray}
By writing 
\begin{eqnarray}
-\frac{i}{2}\int_{\Sigma} \sqrt{g}\ d^2\xi
g^{\alpha\beta}({\partial}_{\beta}\bar{q}){\gamma}_{\alpha}q
&=& \frac{i}{2}\int_{\Sigma}
{\partial}_{\beta}(\sqrt{g}g^{\alpha\beta})\bar{q}{\gamma}
{\alpha}q\ d^2\xi \nonumber \\
 &+& \frac{i}{2}\int_{\Sigma}\sqrt{g}\ d^2\xi
g^{\alpha\beta}\bar{q}({\partial}_{\beta}{\gamma}_{\alpha}q 
+ \frac{i}{2}\int_{\Sigma}\ d^2\xi
g^{\alpha\beta}\bar{q}{\gamma}_{\alpha}{\partial}_{\beta}q
\nonumber \\
&-& \frac{i}{2}\int_{\Sigma} d^2\xi {\partial}_{\beta}\{
\sqrt{g}g^{\alpha\beta}\bar{q}{\gamma}_{\alpha}q\} \nonumber 
\end{eqnarray}
and using
\begin{eqnarray}
{\partial}_{\beta}{\gamma}_{\alpha} &=&
{\partial}_{\beta\alpha}X^{\mu}{\gamma}_{\mu}\ =\
{\Gamma}^{\delta}_{\alpha\beta}{\partial}_{\delta}X^{\mu}
{\gamma}_{\mu} + H^{\mu}_{\alpha\beta}{\gamma}_{\mu},
\nonumber
\end{eqnarray}
where we have made use of the Gauss equation for immersion,
viz., 
\begin{eqnarray}
{\partial}_{\beta}{\partial}_{\alpha}X^{\mu} &=&
{\Gamma}^{\delta}_{\beta\alpha}{\partial}_{\delta}X^{\mu} +
H^{\mu}_{\beta\alpha}
\end{eqnarray}
with the affine connections ${\Gamma}^{\delta}_{\beta\alpha}$
determined by $g_{\alpha\beta}$ and 
\begin{eqnarray}
H^{\mu}_{\alpha\beta} &=& H^i_{\alpha\beta} N^{i\mu},
\end{eqnarray}
where $H^i_{\alpha\beta}$'s are components of the {\it second
fundamental} form along the two normals $N^{i\mu}$ ($i$\ =\
1,2) to the surface $\Sigma$, Eqn.2 becomes 
\begin{eqnarray}
\int {\cal{L}}_D d^2\xi &=& \int_{\Sigma}\sqrt{g}\ d^2\xi i
g^{\alpha\beta}\bar{q}\{ {\gamma}_{\alpha}{\partial}_{\beta} +
\frac{1}{2}H^i_{\alpha\beta}N^{i\mu}{\gamma}_{\mu}\}q,
\end{eqnarray} 
where we have used ${\nabla}_{\beta}(\sqrt{g}
g^{\alpha\beta})\ =\ 0$, ${\nabla}_{\beta}$ being the
covariant derivative on $\Sigma$. We introduce the pull-back
of ${\gamma}_{\mu}$ onto the normal frame by
\begin{eqnarray}
n_i &=& N^{\mu}_i {\gamma}_{\mu},
\end{eqnarray}
and it follows from $\{{\gamma}_{\mu},{\gamma}_{\nu}\}\ =\
2{\delta}_{\mu\nu}$ that
\begin{eqnarray}
\{n_i, n_j\} &=& 2{\delta}_{ij} \nonumber \\
\{n_i, {\gamma}_{\alpha}\} &=& 0.
\end{eqnarray}  
Then Eqn.7 becomes
\begin{eqnarray}
\int {\cal{L}}_D d^2\xi &=& \int_{\Sigma} \sqrt{g}\ d^2\xi i
g^{\alpha\beta} \bar{q}\{ {\gamma}_{\alpha}{\partial}_{\beta}
+ \frac{1}{2}n_i H^i_{\alpha\beta}\}q.
\end{eqnarray}
In this way, the Dirac fermion couples to the world sheet only
through the extrinsic geometry of the surface $\Sigma$. This
agrees with [7,8] and Sedrakyan and Stora [11]. However,
instead of rewriting (10) using spin(n) operators as in [11],
we use (10) here, along the lines of [7,8].

\vspace{1.0cm}

{\bf 4.} We employ generalized Gauss map ${\cal{G}}$
[12,13,14] to describe the immersion of $\Sigma$ in $R^4$.
Then,
\begin{eqnarray}
{\cal{G}}: \Sigma \rightarrow G_{2,4} &\simeq &
SO(4)/(SO(2)\times SO(2)) \ \simeq \ Q_2,
\end{eqnarray}
where the complex quadric $Q_2$ is taken as a model for the
Grassmannian $G_{2,4}$ [12,13]. Then the tangent vector(s)
${\partial}_zX^{\mu}$ is identified as a popint in $Q_2$, upto
a multiplicative complex function $\psi$, namely,
\begin{eqnarray}
{\partial}_z X^{\mu} & =& \psi {\Phi}^{\mu}, \nonumber \\
{\partial}_{\bar{z}}X^{\mu} &=& \bar{\psi}\bar{\Phi}^{\mu}.
\end{eqnarray}
In (12), ${\Phi}^{\mu}{\Phi}^{\mu}\ =\ 0$. This drives the
induced metric $g_{\alpha\beta}$ to be in the conformal gauge,
\begin{eqnarray}
g_{zz} & = & g_{\bar{z}\bar{z}}\ =\ 0, \nonumber \\
g_{z\bar{z}} &=& {\mid \psi\mid}^2 {\mid \Phi \mid}^2,
\nonumber \\
\sqrt{g} &=& {\mid \psi \mid}^2 {\mid \Phi \mid}^2, \nonumber
\\
g^{z\bar{z}} &=& 1/({\mid \psi \mid}^2{\mid \Phi \mid}^2),
\end{eqnarray}
where ${\mid \Phi \mid}^2 \ =\ {\Phi}^{\mu}\bar{\Phi}^{\mu}$.
Using (13), the Dirac action (10) is written as
\begin{eqnarray}
\int {\cal{G}}_D d^2\xi &=&   
  \int_{\Sigma} \sqrt{g}\ (dz\wedge
d\bar{z} \frac{i}{2}) \nonumber \\  
 & & \bar{q}\{ \frac{1}{\sqrt{g}}\left(
{\partial}_zX^{\mu}{\partial}_{\bar{z}} +
{\partial}_{\bar{z}}X^{\mu}{\partial}_z\right)+H^iN^{i\mu}\}
{\gamma}_{\mu} q,
\end{eqnarray}
where (3) and (8) have been used and $H^i\ =\
\frac{1}{2}g^{\alpha\beta}H^i_{\alpha\beta}$. The Dirac
operator is of the form
\begin{eqnarray}
& {\gamma}_{\mu}\left( D^{\mu} + A^{\mu}\right),
\end{eqnarray}
where
\begin{eqnarray}
D^{\mu} &=& \frac{1}{\sqrt{g}}\{
{\partial}_zX^{\mu}{\partial}_{\bar{z}} +
{\partial}_{\bar{z}}X^{\mu}{\partial}_z\}, \nonumber \\
      &  & \nonumber \\
A^{\mu} &=& H^iN^{i\mu}.
\end{eqnarray}  

\vspace{1.0cm}

{\bf 5.} The effective action obtained by integrating over the
fermions is 
\begin{eqnarray}
{\Gamma} &=& \frac{1}{2} Tr \ell n [\gamma \cdot (D + A)]^2,
\end{eqnarray}
where the trace is over the n-dimensional gamma matrices and
over the 2-dimensional surface $\Sigma$. It is seen that
\begin{eqnarray}
[\gamma \cdot (D + A)]^2&=& D_{\mu}D^{\mu} + D_{\mu}A^{\mu} +
A_{\mu}A^{\mu} \nonumber \\
 & & \nonumber \\
&+ &
\frac{1}{2}[{\gamma}_{\mu},{\gamma}_{\nu}]\{\frac{1}{2}
[D^{\mu}, D^{\nu}] + D^{\mu}A^{\nu} + A^{\mu}D^{\nu}\},
\end{eqnarray}
since $A^{\mu}D_{\mu}\ =\ 0$ on account of ${\partial}_zX^\mu\
N^i_{\mu}\ =\ 0$.  The $D_{\mu}D^{\mu}$ term using (13) is
\begin{eqnarray}
D_{\mu}D^{\mu}&=&
\frac{1}{\sqrt{g}}\left({\partial}_zX^{\mu}{\partial}_{\bar
{z}}+{\partial}_{\bar{z}}X^{\mu}{\partial}_z\right)\left(
{\partial}_zX^{\mu}{\partial}^z+{\partial}_{\bar{z}}X^{\mu}
{\partial}^{\bar{z}}\right) \nonumber \\
&= & \frac{{\mid \psi \mid}^2{\mid \Phi \mid}^2}{\sqrt{g}}
\left( {\partial}_z{\partial}^z + {\partial}_{\bar{z}}
{\partial}^{\bar{z}}+{\Gamma}^z_{zz}{\partial}^z +
{\Gamma}^{\bar{z}}_{\bar{z}\bar{z}}{\partial}^{\bar{z}}\right)
, \nonumber \\
&= & {\nabla}_{\alpha}{\nabla}^{\alpha}.
\end{eqnarray}  

The term $D_{\mu}A^{\mu}$ in (18) will be evaluated using the
Weingarten equation for the normals [15]
\begin{eqnarray}
{\partial}_{\alpha}N^{i\mu} &=&
-H^i_{\alpha\beta}g^{\beta\delta}{\partial}_{\delta}X^{\mu} +
(N^{j\nu}{\partial}_{\alpha}N^{i\nu})N^{j\mu}
\end{eqnarray}
as 
\begin{eqnarray}
D^{\mu}A_{\mu} &=& \frac{H^i}{\sqrt{g}}\left(
{\partial}_zX^{\mu}{\partial}_{\bar{z}}N^i_{\mu}+{\partial}_
{\bar{z}}X^{\mu}{\partial}_zN^i_{\mu}\right) \nonumber \\
&=&
-\frac{H^i}{\sqrt{g}}\left(H^i_{\bar{z}\beta}g^{\beta\delta}
g_{\delta z}+H^i_{z\beta}g^{\beta\delta}g_{\bar{z}\delta}
\right) \nonumber \\
&=&
-\frac{H^i}{\sqrt{g}}\left(H^i_{\bar{z}z}+H^i_{z\bar{z}}
\right), \nonumber
\end{eqnarray}
where repeated use of the relations (13) have been made. Now
from $H^i\ =\ \frac{1}{2}g^{\alpha\beta}H^i_{\alpha\beta}$ and
(13), it follows $H^i\ =\
\frac{1}{2}g^{z\bar{z}}(H^i_{z\bar{z}}+H^i_{\bar{z}z})$ and
so,
\begin{eqnarray}
D^{\mu}A_{\mu} &=& -2H^iH^i \ =\ -2{\mid H\mid}^2.
\end{eqnarray}

The term $A_{\mu}A_{\mu}$ is simply $H^iH^i\ =\ {\mid H
\mid}^2$ and thus,
\begin{eqnarray}
[\gamma \cdot (D + A)]^2 &=&
{\nabla}_{\alpha}{\nabla}^{\alpha} - {\mid H \mid }^2 +
 \frac{1}{4}[{\gamma}_{\mu},{\gamma}_{\nu}]\{[D^{\mu},
D^{\nu}] + (D^{\mu}A^{\nu})\} \nonumber \\  
&=& {\nabla}_{\alpha}{\nabla}^{\alpha} + Y,
\end{eqnarray}
thereby defining $Y$.

\vspace{1.0cm}

{\bf 6.} We follow Hawking [16] for the evaluation of the
determinant using the heat-kernel method in curved space. The
proper-time regularization method will be used to write (17)
as
\begin{eqnarray}
\Gamma &=& -\frac{1}{2} Tr {\int}^{\infty}_{0} \frac{ds}{s}
\exp{-s[\gamma \cdot (D+A)]^2},
\end{eqnarray}
after subtracting the divergent part [16]. In the coincidence
limit [16,17], we have
\begin{eqnarray}
\Gamma &=& {\lim_{\xi \to \xi '}}\ -\frac{1}{2}Tr \int
\int^{\infty}_{0}\frac{ds}{s}<\xi \mid \exp{-s[\gamma \cdot 
(D+A)]^2}\mid \xi '> \sqrt{g}\ d^2\xi. 
\end{eqnarray}  
It is to be noted here that the measure $\sqrt{g}\ d^2\xi$ is
in the {\it coincident} limit and Hawking [16] places
$\sqrt{g}$ in the Minakshisundaram-Seeley coefficients after
the coincident limit taken. The heat-kernel has a short
distance expansion [17] which in view of the fact that the
operator here is 2-dimensional gives,
\begin{eqnarray}
\Gamma &=& -\frac{1}{2}\int \sqrt{g}\ d^2\xi Tr
\int^{\infty}_{0}\frac{ds}{s}\ \frac{1}{4\pi s}
\exp{-\frac{\sigma}{2s}}\{a_0 + a_1 s + a_2 s^2+\cdots\}
\end{eqnarray}
where $\sigma$ is the world function [18] which is half the
{\it square} of the geodesic distance between $\xi$ and $\xi
'$. The terms in (22) other than the Laplacian will be
contained in $a_1$ onwards (see [17]). The first term in (25)
can be integrated to give $\frac{2}{\sigma}$. Denote 
$-\frac{1}{\pi\sigma}$ by $I_0$. The second term after $t\ =\
1/s$ substitution gives $\ell n(\frac{2\mu}{\sigma})a_1$,
where $\mu$ is introduced to make the logarithm dimensionless.
Terms $a_2$ onwards vanish when $\sigma \ \rightarrow \ 0$
.(Such terms involve $\int^{\infty}_0 ds s^n
\exp(-\sigma/(2s))$ with $n\ =\ 0,1,2 $ etc, and they are
$\frac{{\sigma}^{n+1}}{2^{n+1}}\Gamma(-n-1)$.). Then (25)
becomes,
\begin{eqnarray}
\Gamma &=& I_0 \int \sqrt{g}\ d^2\xi + \ell
n(\frac{2\mu}{\sigma})\int \sqrt{g} d^2\xi Tr a_1.
\end{eqnarray}
From the expression (22) and from [17], it follows that
\begin{eqnarray}
Tr a_1 &=& \frac{R}{6} - 4{\mid H \mid}^2,
\end{eqnarray}
where $R$ is the scalar curvature of the surface $\Sigma$.
Thus the effective action, after setting $4\ell
n\frac{2\mu}{\sigma}$ by $I_R$ and $\frac{2}{3}\ell
n(\frac{2\mu}{\sigma})$ by $I_E$, becomes,
\begin{eqnarray}
\Gamma &=& I_0 \int \sqrt{g}\ d^2\xi - I_R\int \sqrt{g}\ {\mid
H \mid}^2 d^2\xi + I_E \int \sqrt{g} R d^2\xi.
\end{eqnarray}

\vspace{1.0cm}

{\bf 7.} Thus we have shown that starting from the Dirac
action (2) on $\Sigma$ for {\it free} fermions, the effective
action obtained by integrating the fermion fields consists of
the Nambu-Goto area term, the extrinsic curvature term and the
Euler characteristic. The constants in front of them are
divergent as $\sigma \ \rightarrow \ 0$. It is  important
to see that the first term in (29) which is the NG action has
divergent constant $I_0\ (=-\frac{1}{\pi\sigma})$. Since
$\sigma$ is the half the {\it square} of the geodesic distance
between the coincident points $\xi$ and $\xi '$, ${\sigma}$ 
has the dimensions of $(length)^2$ and so $I_0$ has {\it
$(length)^{-2}$ } dimensions. In string theory [19], the
coeffcient in front of the NG term {\it indeed} has the
dimensions of  $(length)^{-2}$ (in $\hbar$ \ =\ c\ =\ 1
units, used here as well). 
If one starts with a string action , then 
  $I_0$ can be legitimately
absorbed in the string tension. The coefficient in front of
the second term in (29), the extrinsic curvature action, is
dimensionless and can be absorbed in the stiffness parameter.
This demonstration gives a legtimacy for introducing the
extrinsic curvature term in the string action. The importance
of the extrinsic curvature term in QCD strings, in providing
smoothness for the world sheet in lattice calculations [20],
in obtaining physical effects [21], providing the correct high
temperature behaviour [5] and indicating confinement of QCD
flux tubes [4] are well known.  
   
\vspace{1.0cm}

One of the authors (R.P) acknowledges with pleasure
G.Rajasekaran, R.Anishetty, H.Sharatchandra and G.S.Date for
useful discussions.

\vspace{1.0cm}

{\noindent{\bf References}}

\begin{enumerate}

\item A.M.Polyakov, Nucl.Phys. {\bf B268} (1986) 406; \\
      Mod.Phys.Lett. {\bf A2} (1987) 893; \\
      {\it {Gauge Fields and Strings}},Harwood Academic, \\
      Chur.Switzerland. 1987.

\item H.Kleinert, Phys.Lett. {\bf B174} (1986) 335; Phys.Rev.
\\
      Lett. {\bf 58} (1987) 1915; Phys.Lett. {\bf B189} \\
      (1987) 187.

\item K.S.Viswanathan, R.Parthasarathy and D.Kay, Ann.Phys.
\\
      (N.Y) {\bf 206} (1991) 257. 

\item K.S.Viswanathan and R.Parthasarathy, Phys.Rev. \\
      {\bf D51} (1995) 5830.

\item H.Kleinert and A.M.Chervyakov, Phys.Lett. {\bf B381}
      (1996) 286.

\item J.Polchinski, Phys.Rev.Lett., {\bf 68} (1992) 1267.

\item P.B.Wiegmann, Phys.Lett. {\bf B323} (1989) 311.

\item P.B.Wiegmann, Phys.Lett. {\bf B323} (1989) 330.

\item A.M.Polyakov, Nucl.Phys. {\bf B486} (1997) 23.

\item G.Rajasekaran and V.Srinivasan, Prmana. {\bf 10} (1978)
      33.

\item A.G.Sedrakian and R.Stora, Phys.Lett. {\bf B188} (1987)
      442.

\item D.A.Hoffman and R.Osserman, J.Diff.Geometry. {\bf 18}
      (1983) 733.  

\item D.A.Hoffman and R.Osserman, Proc.London.Math.Soc., (3)
      {\bf 50} (1985) 21.

\item R.Parthasarathy and K.S.Viswanathan, Int.J.Mod.Phys.
      {\bf A7} (1992) 317. 

\item L.P.Eisenhart, {\it Riemannian Geometry}, Princeton \\ 
      Univ.Press., Princeton, 1966.

\item S.W.Hawking, Comm.Math.Phys. {\bf 55} (1977) 123.

\item B.de Witt, Phys.Rep. {\bf C19} (1975) 296. \\
      V.N.Romonov and A.S.Schwarz, Teor.Mate.Fizika. {\bf 41}
(1979) 190.

\item J.L.Synge, {\it Relativity, The General Theory}, \\
      Amsterdam, 1960.

\item M.B.Green, J.H.Schwarz and E.Witten, \\
      {\it Super string theory, Vol.I,II}, Cambridge Univ. \\
      Press. Cambridge. 1987.

\item V.A.Kazakov, Phys.Lett. {\bf B128} (1983) 316. \\
      I.K.Kostov,  Phys.Lett. {\bf B138} (1984) 191. \\
      K.H.O'Brien and J.B.Zuber, Nucl.Phys. {\bf B253} (1985)
621.

\item W.Helfrich, Z.Naturforch, {\bf C28} (1973) 693. \\
      T.L.Curtright, G.I.Ghandour, C.B.Thorn and C.K.Zachos,
\\
      Phys.Rev.Lett. {\bf 57} (1986) 799. \\
      T.L.Curtright, G.I.Ghandour and C.K.Zachos, Phys.Rev.
      {\bf D34} (1986) 3811. \\
      K.S.Viswanathan and X.Zhou, Int.J.Mod.Phys. {\bf A3}
(1988) 2195.

\end{enumerate}  
\end{document}